\documentclass[twoside]{IEEEtran}

\usepackage{ifpdf}

\usepackage[noadjust]{cite}

\usepackage[pdftex]{graphicx}

\usepackage[cmex10]{amsmath}
\usepackage[export]{adjustbox}

\usepackage[caption=false,font=footnotesize]{subfig}

\usepackage{xcolor}

\usepackage[english]{babel}
\usepackage[utf8]{inputenc}
\usepackage[T1]{fontenc}

\usepackage[scaled=0.91]{helvet}

\usepackage[cmintegrals,varvw,]{newtxmath}
\usepackage{bm}
\usepackage{siunitx}

\IEEEeqnarraydefcolsep{0}{\leftmargini}
\newcommand*\mat[1]{\textsf{\textit{\textbf#1}}}

\renewcommand*\vec[1]{\textsf{\textit{\textbf#1}}}

\newcommand*\inC[1]{\in\mathbb{C}^{#1}}
\newcommand*\inR[1]{\in\mathbb{R}^{#1}}
\newcommand*\inN[1]{\in\mathbb{N}^{#1}}
\newcommand*\T{^\mathrm{T}}
\newcommand*\adj{^\mathrm{H}}
\newcommand*\inv{^{-1}}
\newcommand*\pinv{^{+}}

\newcommand*\vphinv{^{\vphantom{-1}}}

\newcommand*\mj{\mathrm{j}}
\newcommand*\me{\mathrm{e}}
\newcommand*\diag{\mathrm{diag}}
\newcommand*\DoF{\mathrm{rk}\,\mat A}

\newcommand{\norm}[1]{\left\lVert #1 \right\rVert}

\usepackage[acronym,shortcuts, nonumberlist]{glossaries}
\newacronym{aut}{AUT}{antenna under test}
\newacronym{ff}{FF}{far-field}
\newacronym{nf}{NF}{near-field}
\newacronym{nffft}{NFFFT}{\ac{nf} \ac{ff} transformation}
\newacronym{fiafta}{FIAFTA}{Fast Irregular Antenna Field Transformation Algorithm}
\newacronym{snr}{SNR}{signal-to-noise ratio}
\newacronym{rmse}{RMSE}{root mean square error}
\newacronym{nwa}{VNA}{vector network analyzer}
\newacronym{dof}{DoFs}{degrees of freedom}
\newacronym{svd}{SVD}{singular value decomposition}
\newacronym{rd}{RD}{reconstruction deviation}
\usepackage{blindtext}

\begin{document}
\title{Phase Retrieval for Partially Coherent Observations}

\author{Jonas Kornprobst, \IEEEmembership{Graduate Student Member, IEEE}, Alexander Paulus, \IEEEmembership{Graduate Student Member, IEEE}, \\Josef Knapp, \IEEEmembership{Graduate Student Member, IEEE}, and Thomas F. Eibert, \IEEEmembership{Senior Member, IEEE}%
\thanks{The authors are with the Chair of High-Frequency Engineering, Department of Electrical and Computer Engineering, Technical University  of  Munich,  80290  Munich, Germany (e-mail:  j.kornprobst@tum.de, hft@ei.tum.de).}%
\thanks{This work was supported by the German Federal Ministry for Economic Affairs and Energy under Grant 50RK1923.}%
}

\markboth{IEEE Transactions on Signal Processing}%
{Korprobst \MakeLowercase{\textit{et al.}}: Phase Retrieval for Partially Coherent Observations}


\maketitle

\begin{abstract}
Phase retrieval is in general a non-convex and non-linear task and the corresponding algorithms struggle with the issue of local minima.
We consider the case where the measurement samples within typically very small and disconnected subsets are coherently linked to each other\,---\,which is a reasonable assumption for our objective of antenna measurements.
Two classes of measurement setups are discussed which can provide this kind of extra information: multi-probe systems and holographic measurements with multiple reference signals.
We propose several formulations of the corresponding phase retrieval problem. 
The simplest of these formulations poses a linear system of equations similar to an eigenvalue problem where a unique non-trivial null-space vector needs to be found. 
Accurate phase reconstruction for partially coherent observations is, thus, possible by a reliable solution process and with judgment of the solution quality. 
Under ideal, noise-free conditions, the required sampling density is less than two times the number of unknowns. 
Noise and other observation errors increase this value slightly.
Simulations for Gaussian random matrices and for antenna measurement scenarios demonstrate that reliable phase reconstruction is possible with the presented approach.
\end{abstract}

\begin{IEEEkeywords}
linear phase retrieval, non-convex non-linear cost function minimization, phaseless/magnitude-only near-field far-field transformation, equivalent source reconstruction.
\end{IEEEkeywords}

\section{Introduction}

\IEEEPARstart{P}{hase retrieval} is a type of inverse problem arising in many research fields, including optics~\cite{shechtman_phase_2015,gerchberg_practical_1972}, X-ray crystallography~\cite{Miao.2012, Pfeiffer.2006}, high-frequency engineering~\cite{Isernia.1996,Yaccarino.1999,Paulus.2017b,Knapp.2019}, transmission electron microscopy~\cite{Coene.1992, Faulkner.2004}, coherent diffraction imaging~\cite{GuizarSicairos.2008,CANDES2015,Bacca.2019} and ptychography~\cite{iwen2016fast,Ramos.2019,sissouno2019direct}.
From a mathematical point of view, the behavior of phase retrieval algorithms is often studied for abstract scenarios, e.g., Gaussian random matrices, to develop new solution strategies, since this allows to (statistically) analyze the convergence behavior of the algorithms~\cite{Candes.2013,candes_phase_2015a,candes_phase_2015,Netrapalli.2015,grohs2019stable,iwen2019lower,cheng2020stable,grohs2020phase}.%

For some other than random Gaussian matrices\,---\,e.g., Fourier-base magnitude-only measurements\,---,\,phase retrieval can not be proven to work with absolute certainty, or absolute certainty can only be attained for unrealistic measurement accuracies and with quadratic oversampling rates~\cite{Knapp.2019}. 
Hence, a rather popular strategy is to integrate additional information into the problem formulation, going beyond magnitude-only measurements. 
If feasible, one may change the observation kernel (masking~\cite{Pohl.2014}, exploitation of multiple measurement distances~\cite{Schmidt.2009, Yaccarino.1999,Isernia.1996}) or enforce sparsity~\cite{Jaganathan.2017,Pauwels.2018,Qiu.2017,Wang.2018c,Baechler.2019}.
Another way is to relax the original assumption of magnitude-only observations to some extent.

In this work, we consider the case where (possibly small) subsets of the total observations are captured coherently.
In antenna measurements, this can be achieved with special multi-probe or multi-frequency measurements~\cite{Costanzo.2001,Costanzo.2001b,Costanzo.2005,Costanzo.2008,Paulus.2017b,Sanchez.2020,paulus_2020_eucap20,Knapp2020}. 
In optics, measurements with specialized masks containing structured modulations can be seen as multi-probe measurements~\cite{Pohl.2015}.
A similar problem arises in holography for the stitching of holographic images, where the relation between reference signal (from a reference antenna or beam) and measurement signal is changed.
Examples of holographic antenna measurement techniques are found in~\cite{Junkin.2000,Castaldi.2000,Laviada.2013,LaviadaMartinez.2014,arboleya2015phaseless,Arboleya.2018,tenasanchez2018evaluation,Berlt.2020} and of optical holography in~\cite{Gabor.1949,leith1962reconstructed,barmherzig2019holographic}.
Angular synchronization~\cite{viswanathan2015fast,gao2019multi,iwen2020phase} or phase unwrapping~\cite{goldstein1988satellite,ghiglia1998two} are similar but different.

The phase differences obtained from such measurements are typically disconnected and many phase differences needed for straightforward phase reconstruction by concatenation are missing. 
Thus, the determination of the absolute phases at all the measurement locations remains a non-trivial phase retrieval problem. 
In order to solve the phase-retrieval problem by the concatenation of phase differences, the observations need to be collected in an appropriate manner~\cite{Costanzo.2001b}. 
Due to  practical restrictions related to positioning accuracy and error concatenation, such an approach is only of limited utility.

We do not restrict our investigations to the special case of \ac{nf} antenna measurements and propose several  general formulations of the magnitude-only inverse problem with partially coherent observations together with related solution strategies. With sufficiently oversampled observations and suitable forward operator properties, we show that the phase retrieval problem even can be formulated as a linear system of equations which contains the observation data in a non-linear manner.
The formulations are applicable to any kind of phase retrieval where partially coherent observations are available.
The great advantage is that the search for the global minimum\,---\,even under the influence of measurement errors\,---\,becomes feasible.
Phase retrieval results for synthetic data, noisy and ideal, demonstrate that a solution as close as possible to the true one can be reconstructed with certainty, once the corresponding sampling limit is reached\,---\,given that mild conditions on the forward operator of the inverse problem are fulfilled.

\begin{figure*}
\centering
\includegraphics[width=\linewidth]{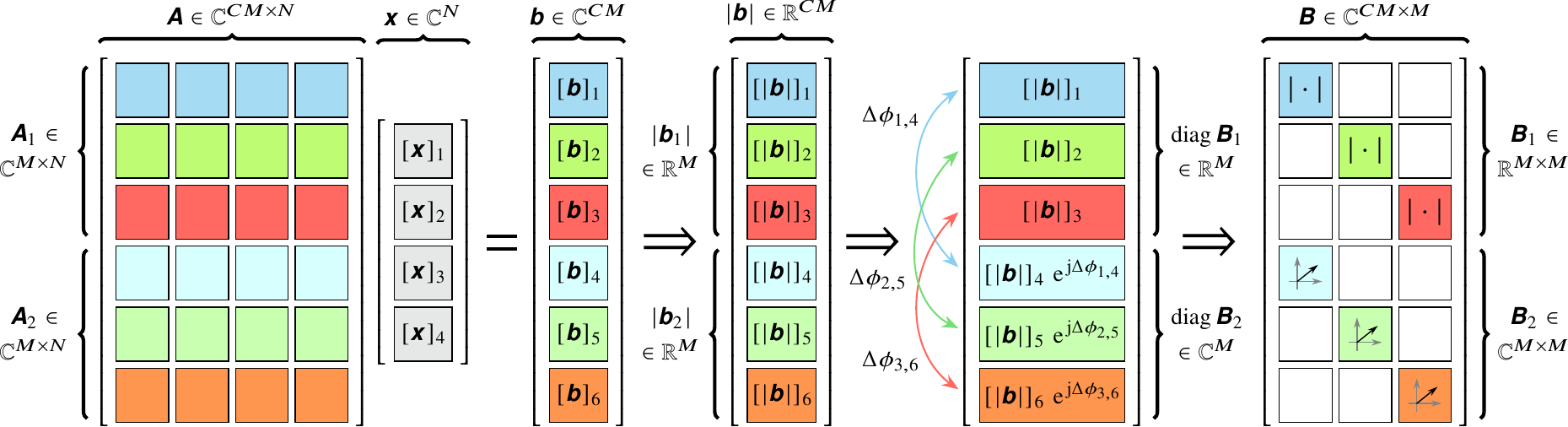}
\caption{An illustration of the ordering of partially coherent observations in the observation vector and observation matrices for the case $N=4$, $M=3$ and $C=2$. From left to right: 
The complex matrix-vector product $\mat A\,\vec x=\vec b$, the magnitude-only observations $|\vec b|$ which are split into the sub-vectors $|\vec b_1|$ and $|\vec b_2|$, the unnamed \emph{“partially-coherent”} observation vector with entry-wise coherence between the first and second halves of the observations, which are employed as the diagonal entries of the diagonal matrices $\mat B_1$ and $\mat B_2$, and, eventually, the \emph{“partially-coherent”} observation matrix $\mat B = [\mat B_1~\mat B_2]\T$.\label{fig:illustration}}
\end{figure*}

The paper is structured as follows. 
Section II introduces the magnitude-only inverse problem. 
Several formulations for phase retrieval with partially coherent observations are proposed in Section III. 
Section IV demonstrates the applicability to random complex matrices and illustrates the limitations of all variants.  
Furthermore, we investigate a possible way of how to incorporate the method into \acp{nffft} with special multi-element probes.

\section{Phase Retrieval\,---\,Problem Statement}
 
\subsection{The Complex Inverse Problem}
In order to introduce the phase retrieval problem statement, we start with a discrete inverse problem $\mat A \,\vec x = \vec b$ written as a least-squares optimization problem
\begin{equation}
\min_{\vec x}~\big\lVert\mat A \,\vec x - \vec b\,\big\rVert_2\label{eq:std_w_phase}
\end{equation}
for retrieving the unknown column vector $\vec x \in \mathbb{C}^N$ from the observation column vector $\vec b \inC{M}$.
The relation between observations and unknowns is established by the forward-operator matrix $\mat A \inC{M\times N}$ with $\mathrm{rk}\,\mat A\le\min(N,M)$. 
We consider $\mathrm{rk}\,\mat A$ as the number of \ac{dof} of the inverse problem. 
We restrict our theoretical investigations to the case of a uniquely defined solution $\vec x$, with  $M\ge\mathrm{rk}\,\mat A=N$.\!\footnote{For reasonably small problems, a rank-revealing decomposition can be employed to reduce the singular case (with $N,M>\DoF$ and $N\ne M$) to the case discussed here.  
This may not be feasible for large $N$ or $M$. }

\subsection{The Classical Magnitude-Only Inverse Problem}
The phase-retrieval problem
\begin{equation}
\min_{\vec x}~\big\lVert|\mat A \,\vec x | - |\vec b|\big\rVert_2\label{eq:std_wo_phase}
\end{equation}
enforces magnitude-only equality between the reconstruction $\mat A \vec x$ and the observations $|\vec b|\inR{M}$, where $|\cdot|$ is the element-wise absolute-value operator.
An alternative formulation of the same problem reads
\begin{IEEEeqnarray}{rCl}
\min_{\bm \phi,\vec x} &&  \big\lVert\mat A \,\vec x  - \diag(\bm\phi)|\vec b|\big\rVert_2\nonumber\\
\mathrm{s.\,t.}       &~~& |[\bm\phi]_m|=1 \quad \mathrm{for~}m\inN{[1,M]}\,,
\label{eq:01}
\end{IEEEeqnarray}
where $\mathbb{N}^{[1,M]}$ represents the natural numbers $\{1,2,\dots,M\}$ and $\diag(\cdot)$ creates a diagonal matrix from a vector (or a vector from the diagonal of a matrix).
An additional unknown vector, the phase vector $\bm\phi\inC{M}$ with the $m$th entry
\begin{equation}
[\bm\phi]_m=\mathrm{e}^{\,\mathrm{j}\phi_m}
\label{eq:phase_vec}
\end{equation}
has been introduced in~\eqref{eq:01}.

\section{The Magnitude-Only Problem with Partially Coherent Observations}
\subsection{Basic Assumptions}
Let us assume that two specific observations\,--\,the $m$th and $k$th ones\,--\,are measured coherently. 
Then, the phase difference between these two observations is known. 
According to~\eqref{eq:phase_vec}, we are able to introduce the additional constraint
\begin{equation}
[\bm\phi]_k/[\bm\phi]_m=\me^{\,\mj(\phi_k-\phi_m)}=\me^{\,\mj\Delta\phi_{k,m}}
\label{eq:PD_km}
\end{equation}
including the observed quantity $\Delta\phi_{k,m}$.
The observed phase difference $\Delta\phi_{k,m}$ is only one out of many, where most may remain unknown. 
Hence, the problem has to be reformulated with all these remaining unknown phase differences in mind.

In order to study the effect of such partially coherent observations, we constrain the way of how these observations are taken.
A special observation probe shall be able to capture $C$ independent observations coherently whenever it performs a measurement.%
\!\footnote{%
This restriction is not necessary in order to benefit from the proposed phase-retrieval method, but it helps to simplify the notation and to predict at which oversampling ratio reliable phase retrieval is possible. 
In some kinds of more general measurements, $C$ may change rather arbitrarily from one observation to another and the derived formulations do also hold in that case. 
However, the notation and, more importantly, the derivation founded on this notation, is simplified considerably by the regular structure of the observations.
}
For an illustrating example, see Fig.~\ref{fig:illustration}, where $N=4$ unkowns and $M=3$ sets of observations taken by a special probe with $C=2$ elements are considered.
The three phase differences $\me^{\,\mj\Delta\phi_{1,4}}$, $\me^{\,\mj\Delta\phi_{2,5}}$, and $\me^{\,\mj\Delta\phi_{3,6}}$ within three distinct pairs of measurement samples are observed in addition to the magnitude-only measurement samples.
With three of these observation pairs\,---\,i.e., $M=3$\,---, we observe $CM=6$ magnitude samples but only $(C-1)M=3$ phase differences. 
$M=3$ phase unknowns remain to be found; this is a compromise between retrieving just a single global phase (fully coherent measurements) and retrieving all $CM=6$ phases (not coherent at all).

\emph{Phase retrieval for partially coherent observations} is a simpler task than the general phase retrieval problem since additional information is available.
Nevertheless, the algorithms found in literature, which tackle this particular problem, are limited to solving non-convex non-linear minimization problems~\cite{Paulus.2017b,Sanchez.2020} or they utilize restrictive\,---\,\emph{unrealistic or even unfeasible}\,---\,sampling strategies~\cite{Costanzo.2001b,Sanchez.2020,Pohl.2017}.

In the general case, the forward matrix is $\mat A \inC{CM\times N}$, the magnitude-only observation vector is~$|\vec b |\inR{CM}$, and the phase unknowns vector is $\bm \phi \inC{CM}$.
The inverse problem 
\begin{IEEEeqnarray}{rCll}
\min_{\bm \phi,\vec x} &~~&  \big\lVert\mat A \,\vec x  - \diag(\bm\phi)|\vec b|\big\rVert_2~&\nonumber\\
\mathrm{s.\,t.}        &  & |[\bm\phi]_m|=1 \quad &\mathrm{for~}m\inN{[1,CM]}
\nonumber\\
                       &  & \frac{[\bm\phi]_{m+cM}}{[\bm\phi]_m}=\me^{\,\mj\Delta\phi_{m+cM,m}} ~~ &\mathrm{for~}m\inN{[1,M]}
\nonumber\\
                       &  &                                                           ~~&\mathrm{and~for~}c\inN{[1,C-1]}
\label{eq:proposed_formulation1}
\end{IEEEeqnarray}
is now additionally constrained by the observed phase differences~$\Delta\phi_{m+cM,m}$. 
Alternatively, we can state that the phase differences $\Deltaup \phi_{k,m}$ according to~\eqref{eq:PD_km} are only observed if $\mathrm{mod}(k-m,M)=0$.
These phase differences do \emph{not} carry the same information as the standard complex data, since they are fewer in number and concatenation is not possible\,---\,for instance, the phase differences from the first subset of $C$ observations $m=\{1,M+1,2M+1,\dots,(C-1)M+1\}$ to \emph{any} other observation outside this subset are missing.

Let us emphasize the key aspect that there are now $CM$ magnitude-only observations and $(C-1)M$ coherent and linearly independent phase-difference observations.
Simple concatenation of the phase differences is not feasible as $M-1$ phase differences plus a global phase are still unknown.
Augmenting the phase retrieval problem~\eqref{eq:01} in~\eqref{eq:proposed_formulation1}, we expect the minimum number of observations for successful reconstruction in the range $N\le CM\le4N$, with the approximate empirical upper bound $4N$ for standard phase retrieval~\cite{candes_phase_2015}.

\subsection{Possible Phase-Difference Measurement Techniques}

\begin{figure}[t]
\centering
\hfill
\subfloat[]{\includegraphics{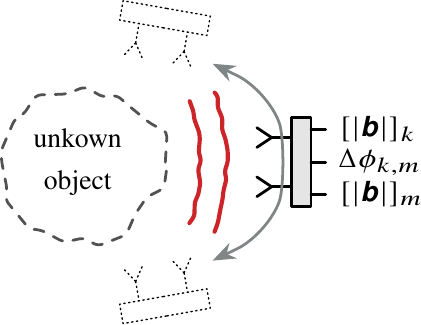}}
\hfill\hfill
\subfloat[]{\includegraphics{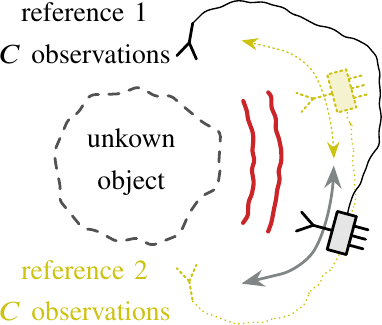}}
\hspace*{\fill}
\caption{Electromagnetic measurement setups for partially coherent observations. (a)~A multi-probe approach, $C=2$. (b)~A holographic approach with two coherent data sets, $M=2$ and $C$ is large.\label{fig:meas_syst}}
\end{figure}

Let us consider two basic types of measurement scenarios: i) multi-probe systems, see Fig.~\ref{fig:meas_syst}(a),
and ii) holographic systems with multiple reference signals, see Fig.~\ref{fig:meas_syst}(b), where reciprocal permutations are possible (the reference may be on the transmitting or receiving side).
Fig.~\ref{fig:meas_syst} focuses on antenna measurement systems; comparable systems in optics have been mentioned in the introduction.

The first case of multi-probe observations seems more challenging since $C$ is typically rather small, e.g., $C=2$ in~\cite{Costanzo.2001,Costanzo.2001b,Costanzo.2005,Costanzo.2008,Paulus.2017b}.
In such a scenario, very small (localized) and possibly unconnected “isles” of coherent observations are taken everywhere on the measurement surface of interest, which is feasible in two ways.
The particular phase differences of interest are observed either directly by multi-channel receivers with shared oscillator signals or via distinct magnitude observations in the form of $[|\vec b|]_k$, $[|\vec b|]_m$, $|[\vec b]_k + [\vec b]_m|$, $|[\vec b]_k + \mj [\vec b]_m|$, which allows to reconstruct the phase differences as
\begin{equation}
\Delta \phi_{k,m}=\mathrm{atan}
\frac{|[\vec b]_k + [\vec b]_m|    ^2-[|\vec b|]_k^2-[|\vec b|]_m^2}
     {|[\vec b]_k + \mj [\vec b]_m|^2-[|\vec b|]_k^2-[|\vec b|]_m^2}
\,.
\end{equation}
Other sets of at least four linear combinations of $[\vec b]_k$ and $[\vec b]_m$ may yield the same reconstructed phase differences.
Essential for the remainder of the paper is only that we can assume the phase differences of the measurement samples within the small “isles” to be known.
Note that, however, the influence of noise differs for different hardware implementations.

In the second case, the relevant task is to stitch holographic sub-images which are, in themselves, fully coherent\,---\,the number of coherent observations $C$ is here typically rather large.
One can think of myriads of variations of such measurement setups, where the constant behavior of the reference signal source has to be known a priori or may even remain unknown but constant. 
One possible antenna measurement scenario, see Fig.~\ref{fig:meas_syst}(b), is to record two coherent sub-images for two installation locations of a receiving reference antenna.

\subsection{Phase Retrieval Formulations with Coherence Constraints}
\subsubsection{A non-linear minimization problem}
The structuring of the $C$ similar blocks is not yet visible in~\eqref{eq:proposed_formulation1}.
Hence, we will explain the various newly introduced variables as shown in Fig.~\ref{fig:illustration}. 
The observation vector~$|\vec b_c| \inR{M}$ represents the $c$th block of the magnitude-only observations, composing the complete  observation vector as, see Fig.~\ref{fig:illustration},
\begin{equation}
|\vec b| =
\begin{bmatrix}
|\vec b_1\T| & \dots & |\vec b_c\T| & \dots & |\vec b_C\T|
\end{bmatrix}\T
\end{equation}
with the corresponding forward-operators $\mat A_c\inC{M\times N}$. Here, the transpose is denoted by $(\cdot)\T$.
Furthermore,  we employ  a reduced phase unknowns vector $\bm\psi\inC{M}$ for the phase unknowns of $|\vec b_1|$ only.\!\footnote{This choice is arbitrary and, due to the flexibility of the phase vector, has no influence on the solution or on the solution process at all.}
The phase differences to the $c$th block are implemented together with the observed magnitudes as a diagonal matrix with entries in the $m$th row and column
\begin{IEEEeqnarray}{lCLl}
[\mat B_1]_{mm}&=&[|\vec b_1|]_m\,,&\mat B_1\inR{M\times M},\\
{}[\mat B_c]_{mm}&=&[|\vec b_c|]_m\,\mathrm{e}^{\,\mathrm{j}\Delta\phi_{m+cM,m}},~&\mat B_c\inC{M\times M},\nonumber\\
        &&&~c\inN{[1,C-1]},\IEEEeqnarraynumspace\label{eq:B_PD_construction}
\end{IEEEeqnarray}%
leading to the overall block-structured matrix, see Fig.~\ref{fig:illustration},
\begin{equation}
\mat B=
\begin{bmatrix} \mat B_1 &\mat B_2&\dots&\mat B_c&\dots&\mat B_C \end{bmatrix}\T\,,~~\mat B\inC{CM\times M}
\end{equation}%
including all observed phase differences and magnitudes. 
The $M$ subgroups of observations are assembled in $\mat B$, where  each column contains $C$ entries including the observed phase differences. Furthermore, the $m$th column has an unknown phase $[\bm\psi]_m$ for its complex entries.

These auxiliary quantities allow to rewrite~\eqref{eq:proposed_formulation1} as
\begin{equation}
\min_{\bm \psi,\vec x}~
\left\lVert
\mat A\,
\vec x - 
\mat B\,
\bm \psi \right\lVert_2~                                                                           ~~ 
\mathrm{s.\,t.}        ~ |[\bm\psi]_m|=1 ~ \mathrm{for~}m\inN{[1,M]}\,.
\label{eq:proposed_formulation2}
\end{equation}

This formulation is already much easier to implement than~\eqref{eq:proposed_formulation1}, since only one non-convex side constraint with reduced dimension  remains.
One simple trick to get rid of this side constraint is to replace the phase unknowns by the reconstructed observations in the manner of
\begin{equation}
\min_{\vec x} ~\,
\left\lVert\,
\begin{bmatrix} |\mat A_1 \vec x | &-&|\vec b_1| \\[0.15cm]
\begin{bmatrix} \mat A_2 \\[-.4\normalbaselineskip] \vdots \\[-.1\normalbaselineskip] \mat A_c \\[-.4\normalbaselineskip] \vdots \\[-.1\normalbaselineskip] \mat A_C \end{bmatrix}
\vec x &-& 
\begin{bmatrix} 
\mat B_2\\[-.4\normalbaselineskip] \vdots \\[-.1\normalbaselineskip] \mat B_c \\[-.4\normalbaselineskip] \vdots \\[-.1\normalbaselineskip] \mat B_C  
 \end{bmatrix} \mat B_1^{-1}\mat A_1\vec x
\end{bmatrix}
\,
  \right\lVert_2~                                                                            
\,.
\label{eq:proposed_formulation2a}
\end{equation}
Employing $\bm\psi=\mat B_1^{-1}\mat A_1\vec x$ instead of $|[\bm\psi]_m|=1$ as in~\eqref{eq:proposed_formulation2} is an alternative restriction. Its implications remain to be studied.

\subsubsection{A linear formulation for the original unknowns}
The non-linear side constraints (dependent on the type of solver, including a yet-to-determine weighting) and the additional $\bm \psi$ unknowns might be obstructive to deal with. 
Hence, we analyze under which circumstances a unique $\bm\psi$ exists\,---\,rendering the magnitude-one side constraint obsolete.

We proceed by looking at the $C$ sub-equations of~\eqref{eq:proposed_formulation2}
\begin{equation}
\mat A_c \vec x = \mat B_c \bm  \psi\,.
\end{equation}
The matrices $\mat B_c$ are diagonal with non-zero entries and, thus, invertible. Solving for $\bm\psi$ yields
\begin{equation}
\bm\psi =  \mat B_1\inv\mat A_1\vphinv \vec x =\hdots=  \mat B_c\inv \mat A_c\vphinv \vec x =\hdots=   \mat B_C\inv\mat A_C\vphinv \vec x \,,
\label{eq:gen_EV_prblm1}
\end{equation}
which resembles a concatenation of generalized (pseudo-) eigenvalue problems~\cite{toh1996calculation,trefethen1997pseudospectra,trefethen1991pseudospectra,boutry2005generalized,chu2006generalized,kressner2014generalized} (with a known eigenvalue of value $1$) and enables us to eliminate the phase unknowns from the problem
by simultaneously solving $C-1$ generalized eigenvalue problems (with $\lambda=1$) in the form of
\begin{equation}
 \mat B_1\inv\mat A_1\vphinv \vec x = \lambda \mat B_c\inv \mat A_c\vphinv \vec x \quad\mathrm{for}~c\inN{[2,C]}
\,.
 \label{eq:one_of_gen_ev_problems}
\end{equation}
Multiplying by the diagonal matrices $\mat B_1\mat B_c$ leads to\begin{equation}
 \mat B_c\mat A_1\vphinv \vec x =  \mat B_1 \mat A_c\vphinv \vec x \quad\mathrm{for}~c\inN{[2,C]},
 \label{eq:lambda=1}
\end{equation}
and allows to recast the overall problem as
\begin{equation}
\mat Q  \,
\vec x =
\begin{bmatrix}
 \mat B_2 \mat A_1 -   \mat B_1 \mat A_2 \\[-.4\normalbaselineskip]
 \vdots\\[-.1\normalbaselineskip]
 \mat B_c \mat A_1 -   \mat B_1 \mat A_c \\[-.4\normalbaselineskip]
 \vdots\\[-.1\normalbaselineskip]
 \mat B_C \mat A_1 -   \mat B_1 \mat A_C
\end{bmatrix}
\vec x
=\vec 0
\label{eq:gen_EV_prblm2}
\end{equation}
with the matrix $\mat Q \inC{M(C-1)\times N}$.

Equation~\eqref{eq:gen_EV_prblm2} is just the linear and homogeneous part of~\eqref{eq:proposed_formulation2a} multiplied by $\mat B_1$ in order to avoid the division by potentially small $\mat B$ entries and without  the magnitude constraint  $|\mat A _1 \vec x|=|\vec b_1|$. 
Having dropped the non-linear part of the constraints implies that not all information is employed in the source reconstruction. This may lead to a somewhat suboptimal solution, but with the benefit of solving a linear system of equations\,---\,implying an improvement in reliability.

The properties of~\eqref{eq:gen_EV_prblm2} are influenced by the properties of~$\mat A_c$. 
We have assumed that $\mathrm{rk}\,\mat A_c=\min\{M,N\}$.
Hence, with a known $\vec x$, $\bm \psi$ follows immediately. 
However, it is still unclear under which circumstances $\vec x$ is unique.

Due to the block subtractions $\mat B_c \mat A_1- \mat B_1 \mat A_c$ in the matrix $\mat Q \inC{M(C-1)\times N}$, the true $\vec x$ has to be in  $\mathrm{ker}\,\mat Q$. 
We deduce a condition for a unique reconstruction: 
There has to be a non-trivial $\mathrm{ker}\,\mat Q$ with $\mathrm{dim}\,\mathrm{ker}\,\mat Q=1$. 
To fulfill this, we recall that $\mat Q$ has $N$ columns.
Hence, for $\mathrm{dim}\,\mathrm{ker}\,\mat Q=1$, it is required that $\mathrm{rk}\,\mat Q=N-1$. 
In order to yield this rank, a necessary but insufficient condition on  the number of rows of $\mat Q$ reads
\begin{equation}
M(C-1)\ge N-1
\,.
\label{eq:SVD_condition}
\end{equation}
A further requirement is that the eigenvalue $\lambda = 1$ in the general eigenvalue problems~\eqref{eq:one_of_gen_ev_problems} is unique, i.e., it is not degenerate. 
For the subtraction of two random Gaussian matrices, the probability of having a degenerate eigenvalue approaches zero.
For other matrices, this becomes more difficult to achieve, see the application example in Section~\ref{sec:res_nffft}. 
If noisy observations are contained in the matrices $\mat B_c$, degenerate or near-degenerate eigenvectors may prevent a reliable reconstruction.
This is further discussed when measurement errors are considered.

\begin{figure}[t]
\centering
\includegraphics{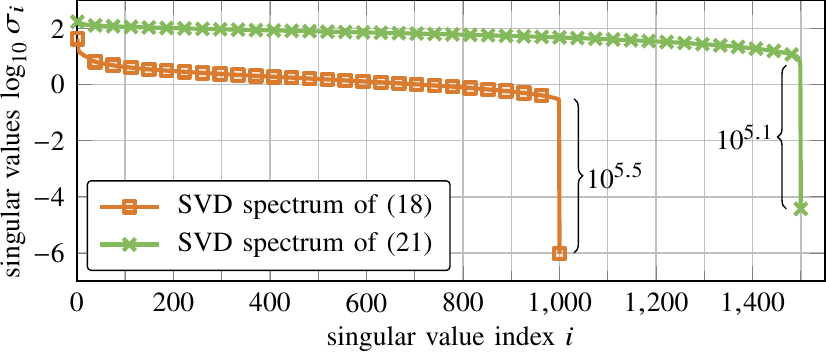}
\caption{\ac{svd} spectra for a random matrix and random right-hand side, with $N=1000$, $M=1500$, $C=2$, and $n=10^{-6}$.\label{fig:svd_null}}
\end{figure}

The task is now to determine the unique non-trivial vector in $\mathrm{ker}\,\mat A_c$. 
We consider the example of a Gaussian random $\mat A$ and $\vec b$ with $N=1000$, $M=1500$, $C=2$, and a noise-to-signal ratio in the observation vector $n=10^{-6}$ according to
\begin{equation}
n = \frac{\lVert\vec b' - \vec b\rVert_2}{\lVert\vec b\rVert_2}\,,\label{eq:noise2signalratio}
\end{equation}
where the primed vector $\vec b'$ contains the noise-contaminated observations.
The spectrum of the \ac{svd} of $\mat Q= \mat U \mat S \mat V\adj$ is shown in Fig.~\ref{fig:svd_null}. %
Exactly one noise-limited singular value with a magnitude of about $10^{-6}$ is observed. 
In a magnitude-ordered \ac{svd}, we refer to this vale as $\sigma_N$.
The corresponding singular vector $\vec{v}_{N}$ solves the phase-retrieval problem to a comparable accuracy level.

The task of finding the non-trivial vector in the null space can be tackled in different ways. 
The obvious one is to perform an \ac{svd} and pick the vector for the smallest singular value, but this is computationally rather expensive. 
From a complexity point of view, we can employ an iterative approach (preferably a Krylov-subspace method, e.g., Arnoldi iterations~\cite{arnoldi1951principle}) to estimate the required \ac{svd} vector and scale the retrieved~$\vec x$ appropriately afterwards. 
Such an iteratively attained solution is also unique once the discussed conditions on $\mat Q$ are fulfilled. 
Furthermore, it does not suffer from local minima, i.e., it is independent from the initial guess.\!\footnote{%
The initial guess is an important concept for non-convex minimization problems. 
Due to the possibility of local minima on the path to the global solution, it matters at which \emph{initially guessed} vector the minimization starts.}
This is explained by the linearity (and, thus, convexity) of the formulation. 
Since the nullity of $\mat Q$ is one, and~\eqref{eq:gen_EV_prblm2} is a homogeneous linear system of equations, even standard solvers for linear systems of equations may be employed if the trivial solution is avoided. 

The formulation offers two possibilities to judge whether the reconstruction was successful.
Firstly, a drop in the SVD spectrum between the smallest and the second smallest singular value should be observable. 
Secondly, the reconstructed phase vector $\bm \psi$ according to~\eqref{eq:gen_EV_prblm1} is required to have entries with constant magnitude.
Unit-magnitude entries of the phase vector are to be created by a suitable scaling. 
If the magnitudes of the vector entries fluctuate a lot, the reconstruction failed.

\subsubsection{A linear formulation for the phase unknowns}
In the step from~\eqref{eq:proposed_formulation2} to~\eqref{eq:proposed_formulation2a},  the phase unknowns were replaced by~$\vec x$. 
However, it is also possible the other way round. If $\bm \psi $ is unique according to~\eqref{eq:SVD_condition}, $\vec x$ may be replaced by the Moore-Penrose pseudo-inverse $\mat A^+$ applied to the “complex” observations $\mat B\,\bm\psi$ yielding again a linear null-space equation
\begin{equation}
\mat R \, 
\bm \psi =
\begin{bmatrix}
\mat A \mat A\pinv \mat B - {\mat B}
\end{bmatrix}
\bm \psi
=\vec 0
\label{eq:nullspace_phases}
\end{equation}
with the matrix $\mat R \inC{CM\times M}$.

In Fig.~\ref{fig:svd_null}, the spectrum of $\mat R=\mat U\mat S\mat V\adj$ is shown for a random matrix with $N=1000$ and $n=10^{-6}$. 
Again, one singular vector $\vec v_M$ of a non-trivial noise-limited null-space (with a singular value $\sigma_M$) appears.
While the null-space seems to have a numerically decreased dynamical range, the reconstruction quality is the same in this particular example. 
Whether one of the two versions is superior to the other is studied in the results sections. 
The main difference to~\eqref{eq:gen_EV_prblm2} is found in the fact that the forward operator appears only in the form of the projector $\mat A \mat A\pinv$, removing most of the influence of the spectrum or null-space of $\mat A$.

After the reconstruction of $\bm\psi$ by~\eqref{eq:nullspace_phases}, the intuitive approach for the solution of the phaseless problem is to solve
\begin{equation}
\mat A \,\vec x = \mat B \,\bm \psi
\label{eq:rec1}
\end{equation}
in a subsequent step. 
This is done in the following unless stated otherwise.
Note, however, that the magnitude-one constraint in the reconstruction of $\bm \psi$ was neglected in order to obtain a linear system of equations. 
Since not all information is considered in the reconstruction process, the retrieved solution may be suboptimal. 
Instead of $\mat B \,\bm \psi$, the complex measurement vector may be reconstructed as
\begin{equation}
\mat A \,\vec x = \mat B \,\mathrm{diag}(|\bm \psi|)^{-1} \bm \psi
\label{eq:rec2}
\end{equation}
enforcing the non-linear magnitude-one constraint.

\subsection{Discussion of the Linearized Reconstruction Algorithms}
\subsubsection{Suboptimality of the Retrieved Solution}

An important question is what happens if the conditions on $\mat Q$ (or, equivalently, $\mat R$) are not met. This might happen if the number of observations is not sufficiently large or if \mbox{(near-)~}degenerate eigenvalues appear due to observation errors. 
Then, the numerically determined nullity $\mathrm{dim}\,\mathrm{ker}\,\mat Q$ is greater than one. 
We still know that the true solution $\vec x\in \mathrm{ker}\,\mat Q$, but \eqref{eq:gen_EV_prblm2} or \eqref{eq:nullspace_phases} alone are not sufficient anymore. 
This offers two strategies.
Either the minimization problem is constraint by choosing only search vectors in $\mathrm{ker}\,\mat Q$, 
or the null-space equation is augmented by additional constraints. 
For instance, the phase vector constraints  $|[\bm\psi]_m|=1$ may be included again.
Another way to get rid of false solutions  in $\mathrm{ker}\, \mat{Q}$ for the homogeneous equation~\eqref{eq:gen_EV_prblm2} is to fix the phase and the magnitude of up to $C$ coherent (and hence complex) observations. 
As part of the following error analysis, a single $i$th entry in $\bm \psi$ is fixed in order to obtain the inhomogeneous and invertible linear system
\begin{equation}
\mat R_{\star} \bm \psi = \vec{u}_{CM+1}\quad\mathrm{with~}
\mat R_{\star} = \begin{bmatrix}
\mat R\T & \vec u_i
\end{bmatrix} \T
\label{eq:noise_Rstar}
\end{equation}
from~\eqref{eq:nullspace_phases}, where $\vec u_i$ refers to the $i$th unit vector.

\subsubsection{Influence of Measurement Errors}
We assume that a noise-contaminated observation vector $\vec b^{\prime} = \vec b +\updelta \vec b$, see~\eqref{eq:noise2signalratio}, leads to perturbations  $\updelta \mat B$ and  $\updelta \mat R$ via a non-linear relation.
From the perturbation theory of linear systems~\cite{Demmel.1997} applied to~\eqref{eq:noise_Rstar}, we state the first-order bound 
\begin{equation}
\frac{\norm{\updelta \bm \psi}}{\norm{\bm \psi}} 
\approx
\frac{\norm{\mat R_{\star}\pinv\updelta \mat R_{\star} \bm \psi}}{\norm{\bm \psi}}
   \le  \norm{\mat R_{\star}\pinv}  \norm{\updelta \mat R_{\star}} 
   \,.
\end{equation}
With $\norm{\updelta \mat R_{\star}} \le \norm{\updelta \mat B}$, we have
\begin{equation}
\frac{\norm{\updelta \bm \psi}}{\norm{\bm \psi}}
   \lesssim  \kappa \frac{\norm{\updelta \mat B} }{\norm{\mat B} }
   \,,
   \label{eq:nsr_bound}
\end{equation}
which holds for any submultiplicative and unitarily invariant matrix norm $\norm{\cdot}$.
Since factoring out $\mat B $ from $\mat R_{\star}\pinv$ is prohibited by the pseudo-inversion, the dependence on the problem sensitivy $\kappa=\norm{\mat R_{\star}\pinv} \norm{\mat B}$ remains.
Relating the bound to ${\norm{\updelta \vec b} }/{\norm{\vec b} }$ is not easy to achieve, since $\norm{\updelta \vec b}\lesssim\norm{\updelta \mat B}$.

\begin{figure}[!tp]
\centering%
\subfloat[]{\includegraphics{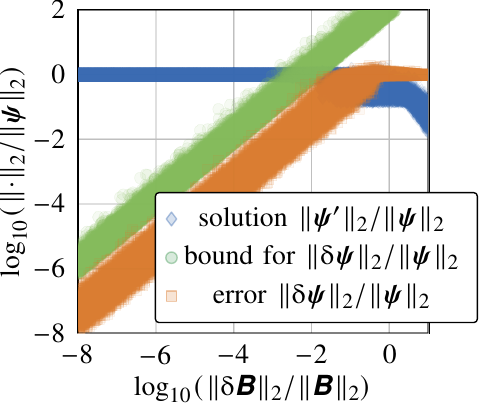}}\hspace*{-0.2cm}
\subfloat[]{\includegraphics{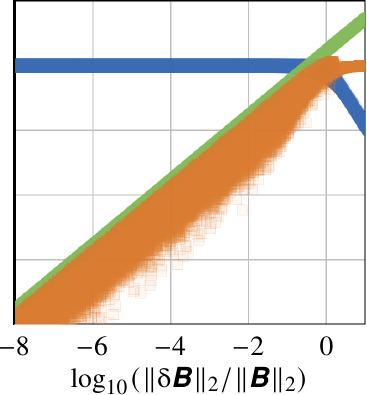}}%
\caption{Simulation results for the first-order bound~\eqref{eq:nsr_bound}, $CM=100$, $N=40$, and $10^5$ random realizations of $\mat A$, $\vec b$, and $\updelta \vec b$.
(a)~$C=2$.
(b)~$C=50$.
\label{fig:nsr_bound}}
\end{figure}%
As observed in Fig.~\ref{fig:nsr_bound}, noise-to-signal ratios above $n\approx1$ lead to a failure of the algorithm since the one-dimensional nullspace of $\mat R$ disappears due to noise.
It is not possible anymore to find a suitable phase vector and, hence, a zero vector from the null space of $\mat R$ is retrieved.
The other way round this means that a zero singular value, which is unique in the noiseless case, does not become degenerate in the presence of small enough noise. 
The error bound proves \eqref{eq:noise_Rstar} to be stable since limited measurement noise has limited impact on the solution vector.
Note that the error is generally lower for large values of $C$ than for smaller values of $C$.
Due to the similarities in the approaches, we expect a similar upper bound to hold true for~\eqref{eq:gen_EV_prblm2}.

\subsubsection{Observations with Zero Magnitude}
At first sight, one might be afraid that measurement samples with zero magnitude cause problems.
Indeed, it is impossible to define a phase difference between two complex numbers with one of them being zero. 
However, a (magnitude-only) observation with zero magnitude constitutes a complex measurement with zero real and imaginary parts.
Since the entries of $\mat B$ as defined in~\eqref{eq:B_PD_construction} contain plain zeros for cases where the phase difference can neither be observed nor defined, no problem arises for the proposed algorithms.
For a complete tuple of $C$ zero observations, 
 we can even rewrite the condition in~\eqref{eq:SVD_condition} as
\begin{align}
CM-\text{rk}\,\mat B\ge N-1\,,
\end{align}
where $\text{rk}\,\mat B$ and the required number of measurements decrease from $M$ by one for each complete tuple of zero-observations.

\subsubsection{On the Computational Complexity}
The computational complexity of the proposed linear formulations is dominated by the inversion of the linear forward operator $\mat A$.
The most challenging case is \eqref{eq:nullspace_phases}. 
If a direct solver is utilized, it is employed twice: for the calculation of the pseudo-inverse $\mat A\pinv$ and for the inversion of $\mat R$. 
If an iterative solver is utilized\,---\,possibly even with an efficient evaluation of $\mat A$\,---,\,it may also be employed in a nested manner for each evaluation of $\mat R$.
In any case, only the prefactor of the computational complexity of a direct or an iterative solver is increased, where this may vary dependent on the formulation.

\subsection{State of the Art: Algorithms for Comparison}
For the standard phase-retrieval problem~\eqref{eq:std_wo_phase}, we consider the algorithms provided by PhasePack~\cite{gerchberg_practical_1972,fienup_phase_1982,Candes.2013,candes_phase_2015,Zhang.2016,chandra_phasepack_2017,dhifallah2018phase,goldstein2018phasemax}.
The only generally applicable method  for incorporating partial knowledge about phase differences, which may be employed to compare it with the proposed algorithm, is found in~\cite{Paulus.2017b}. 
There, the structure of the forward operator is changed, leading to the non-linear non-convex minimization problem
\begin{equation}
\min_{\vec x} ~~ 
\left\lVert~
\left|
\begin{bmatrix} \mat A_1 \\ \mat A_2 \\ \mat A_1 + \mat A_2 \\ \mat A_1 +\mj \mat A_2 \end{bmatrix}
\vec x \,
\right|
-
\begin{bmatrix}|\vec b_1|\\|\vec b_2|\\|\vec b_1 + \vec b_2|\\|\vec b_1 +\mj \vec b_2|\end{bmatrix} \right\lVert_2
\,,\label{eq:PAULUS}
\end{equation}
here given for the case $C=2$.
The phase-difference constraint is included in the cost functional\,---\,and not written as a side constraint.
Of course,~\eqref{eq:PAULUS} can be rewritten for larger $C$. 

\section{Results for Gaussian Random Matrices}
The linear systems of equations are solved as described in Section III. Non-linear minimization problems are solved with the cost function minimizers provided by Matlab~\cite{matlab_version_2019}, where the active-set method is employed for the minimization with equality side constraints. Custom implementations for various solvers capable of handling problems of larger size have also been realized, based on the memory limited $L$-BFGS method~\cite{Liu.1989, Nocedal.2006}, by Broyden, Fletcher, Goldfarb, and Shannon.
The initial guess~$\vec x_0$ for the non-linear solvers \eqref{eq:proposed_formulation1}, \eqref{eq:proposed_formulation2}, \eqref{eq:proposed_formulation2a}, and \eqref{eq:PAULUS} is obtained by a spectral method as~\cite{candes_phase_2015}
\begin{equation}
\vec x_0 = \vec v_{\mathrm{max}} \frac{|\vec b\T| \mat A \,\vec v_{\mathrm{max}}}{\lVert\mat A \,\vec v_{\mathrm{max}}\rVert_2^2}\,,\label{eq:spectral_method}
\end{equation}
with the eigenvector $\vec v_{\mathrm{max}}$ corresponding to the largest eigenvalue of $\mat A\adj \mat B\adj \mat B\mat A$.
The non-linear solvers rely on the initial guess to avoid local minima. The solvers based on the null-space search do not require an initial guess.

\begin{figure}[!tp]
\centering
\subfloat[]{\includegraphics{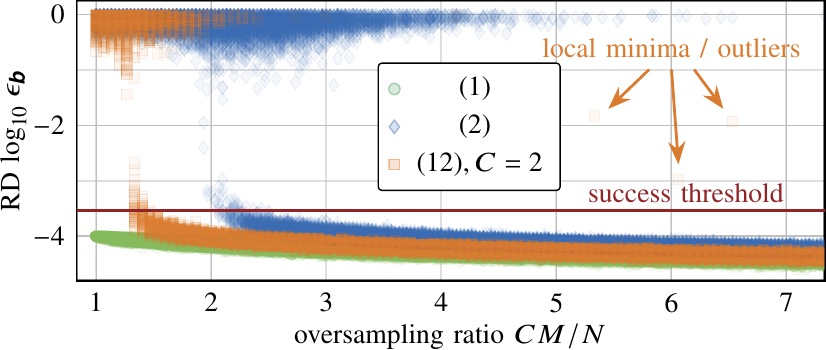}}\\
\subfloat[]{\includegraphics{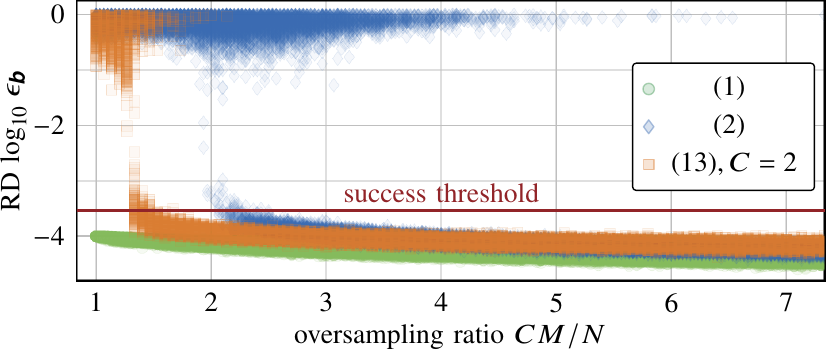}}
\caption{RD of the inverse problem solved for Gaussian random matrices, $N=30$, $C=2$, $n=10^{-4}$. 
Each $CM/N$-ratio with $200$ simulations. 
(a)~Magnitude cost-functional minimization with reduced phase unknowns. 
(b)~Magnitude cost-functional minimization with eliminated phase unknowns. 
\label{fig:scatter_split_1}}
\end{figure}%

\subsection{An Extensive Solver Comparison for $N=30$}
Considering $N=30$,  the phase reconstruction is performed for 200 random picks of $\mat A$, each with a randomly picked true solution $\bm \xi$ and a corresponding right-hand side $\vec b = \mat A\,\bm \xi$.
After the solution process, the true \ac{rd}
\begin{equation}
\epsilon_{\vec b} = {\lVert\mat A \vec x - \mat A\,\bm \xi\rVert_2}/{\lVert\mat A\,\bm \xi\rVert_2}
\end{equation}%
is evaluated, 
where, however, the solution $\vec x$ is obtained for a noise-contaminated vector $\vec b'$ with the noise-to-signal ratio $n$.

For $M\inN{[30,220]}$ and $n=10^{-4}$, the results are shown in Fig.~\ref{fig:scatter_split_1}(a)
for the standard solver with phase~\eqref{eq:std_w_phase}, the standard magnitude-only solver~\eqref{eq:std_wo_phase}, and the proposed solver with a $C=2$ phase-differences side constraint according to~\eqref{eq:proposed_formulation2}.
The scatter plot provides the insight that the fully-coherent complex solver always works and the magnitude-only versions require a certain oversampling for a reliable reconstruction. 
We further observe that the solver with $C=2$ converges with fewer $CM$. 

\begin{figure}[!tp]
\centering
\subfloat[]{\includegraphics{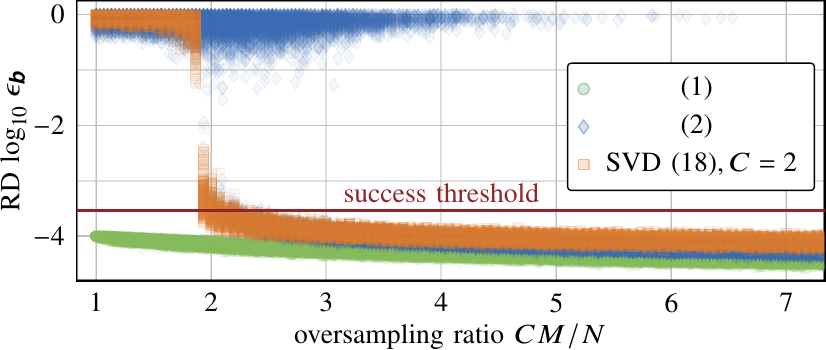}}\\
\subfloat[]{\includegraphics{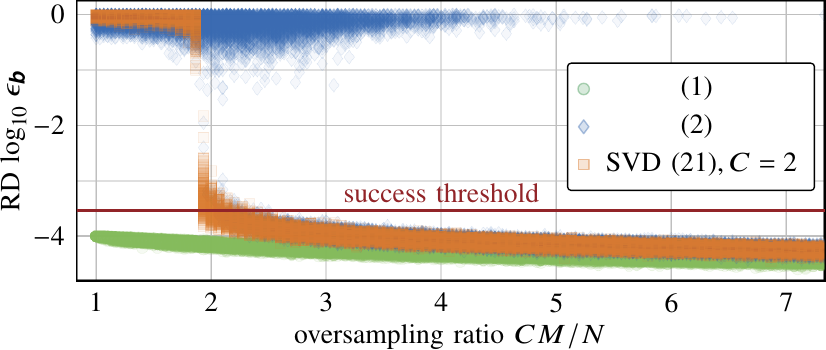}}\\
\subfloat[]{\includegraphics{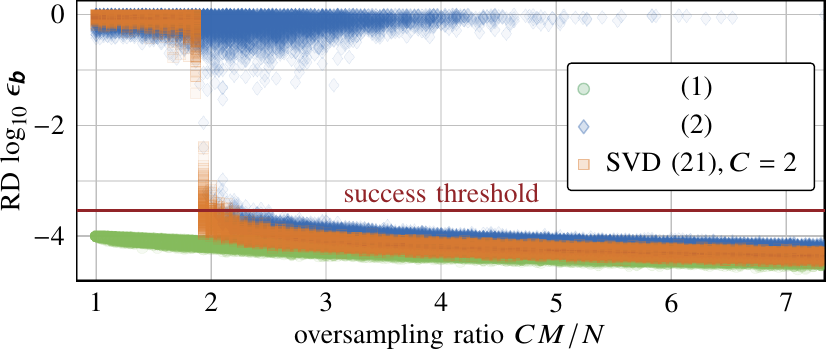}}
\caption{RD of the inverse problem solved for Gaussian random matrices, $N=30$, $C=2$, $n=10^{-4}$. 
Each $CM/N$-ratio with $200$ simulations. 
(a)~First null-space search with SVD. 
(b)~Second null-space search with SVD and fully linear reconstruction~\eqref{eq:rec1}.
(c)~Second null-space search with SVD and enforced magnitude-one constraint~\eqref{eq:rec2}.
\label{fig:scatter_split_2}}
\end{figure}

In Fig.~\ref{fig:scatter_split_1}(b), the fusion of magnitude-minimization and null-space condition~\eqref{eq:proposed_formulation2a} shows a better convergence than~\eqref{eq:proposed_formulation2}, with the main advantage of avoiding to get stuck in local minima for this scenario. 
In Fig.~\ref{fig:scatter_split_2}(a), 
the linear formulation, i.e., the null-space vector search, is included. The convergence behavior is slightly different. Almost exactly at $CM/N=58/30\approx1.93$ as expected, we observe a certain convergence. 
In contrast to Fig.~\ref{fig:scatter_split_1}(a), there are no outliers (i.e., local minima) above $CM/N>2$.
Unfortunately, the limit for successful reconstruction is a bit higher than in the other two cases.
The second null-space equation~\eqref{eq:nullspace_phases} shows a comparable behavior in Fig.~\ref{fig:scatter_split_2}(b) with slightly lower \ac{rd}s.
In Fig.~\ref{fig:scatter_split_2}(c), the magnitude-one constraint for the reconstructed phase unknowns is enforced according to~\eqref{eq:rec2}. 
This seems to improve the RD if the reconstruction is working, i.e., if the success threshold $CM/N\ge58/30$ is fulfilled.

We introduce a threshold of $3n$ with the noise-to-signal ratio $n=10^{-4}$ as defined in~\eqref{eq:noise2signalratio} and call a reconstruction with a RD below this limit \emph{successful} and above \emph{failed}. 
As seen in Figs.~\ref{fig:scatter_split_1} and~\ref{fig:scatter_split_2}, this is a rather demanding definition of a successful reconstruction which excludes three kinds of solutions: global false solutions due to insufficient sampling, wrong solutions due to local minima, and almost acceptable solutions, where, e.g., the solver convergence was too slow.

\begin{figure}[!t]
\centering
\includegraphics{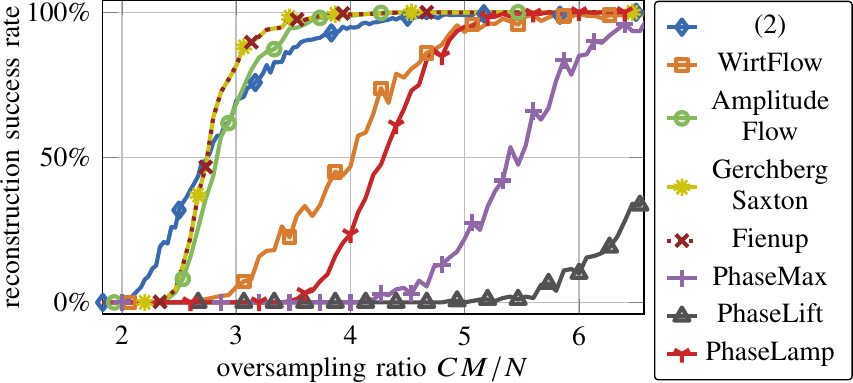}
\caption{Success rate for Gaussian matrices, $N=30$, noise $n=10^{-4}$, and for the comparison $CM/N=M/N$, PhasePack solvers~\cite{chandra_phasepack_2017}.\label{fig:success1a}}
\end{figure}

This allows us to introduce a success rate for the reconstruction. 
In Fig.~\ref{fig:success1a}, the minimization of~\eqref{eq:std_wo_phase} with the $L$-BFGS method is compared to many methods provided by PhasePack~\cite{chandra_phasepack_2017} and the simple cost function minimization is among the best-performing solvers. 
It is a reasonable choice that \eqref{eq:PAULUS} (the comparison method for partially coherent phase retrieval) is minimized  with the L-BFGS method in the following.

\begin{figure}[!tp]
\centering
\subfloat[]{\includegraphics{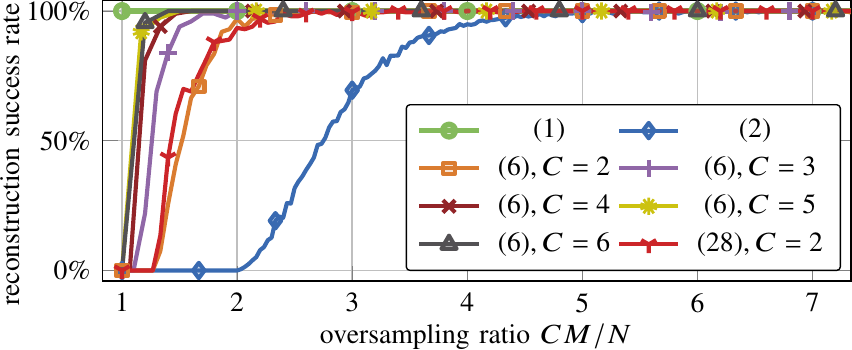}}\\
\subfloat[]{\includegraphics{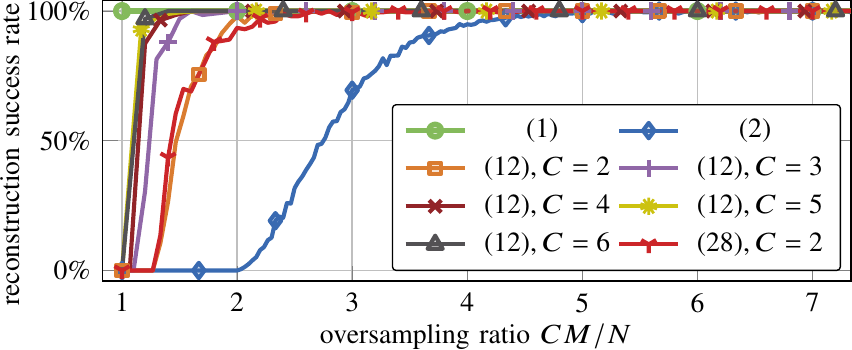}}\\
\subfloat[]{\includegraphics{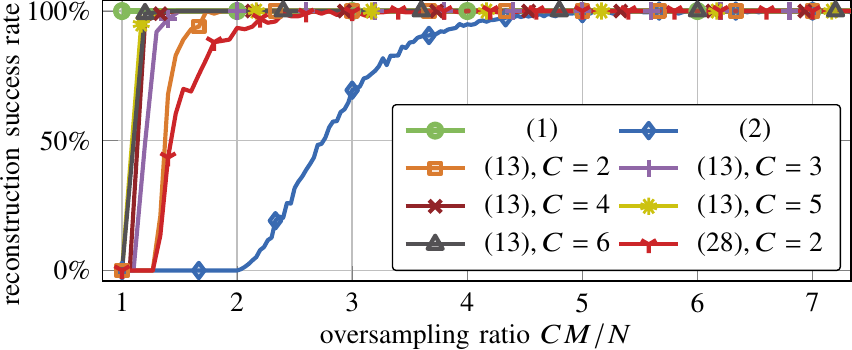}}\\
\subfloat[]{\includegraphics{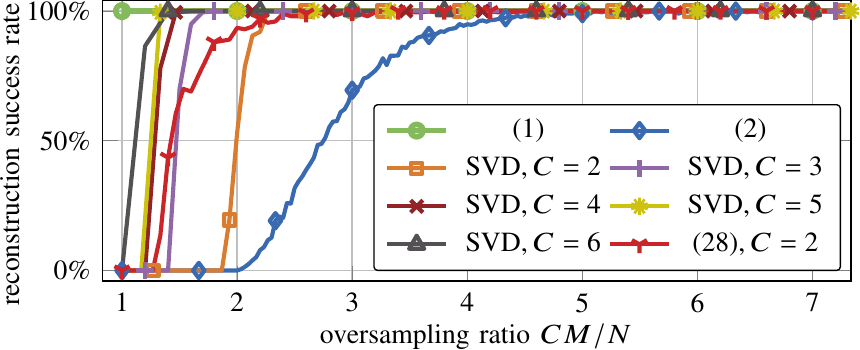}}\\
\subfloat[]{\includegraphics{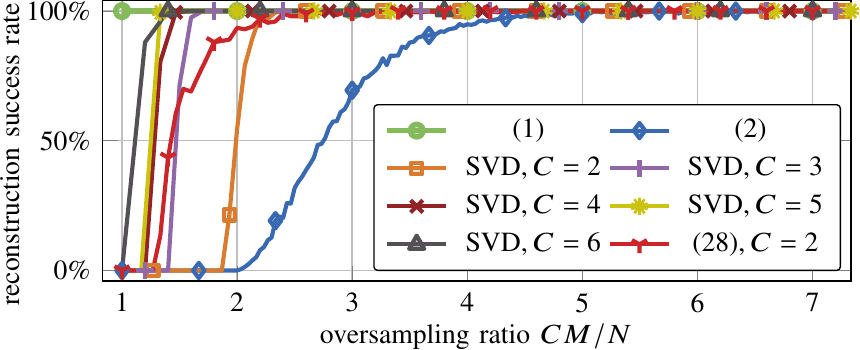}}
\caption{Success rate for Gaussian matrices, $N=30$, noise level $n=10^{-4}$, and for the comparison $CM/N$. 
(a)~Different non-convex solvers~\eqref{eq:proposed_formulation1} including phase unknowns.
(b)~Different non-convex solvers~\eqref{eq:proposed_formulation2} including a reduced number of phase unknowns. 
(c)~Different non-convex solvers~\eqref{eq:proposed_formulation2a}  with eliminated phase unknowns.
(d)~Linearized solvers~\eqref{eq:gen_EV_prblm2} with SVD.
(e)~Linearized solvers~\eqref{eq:nullspace_phases} with SVD.
\label{fig:success1}}
\end{figure}

In Fig.~\ref{fig:success1}(a), the required oversampling ratio $CM/N$ for a high chance of success for the phase-difference solvers~\eqref{eq:proposed_formulation1} moves towards a value of $CM/N=1$ with increasing $C$. 
The same is observed for~\eqref{eq:proposed_formulation2} in Fig.~\ref{fig:success1}(b).
The standard phaseless solver~\eqref{eq:std_wo_phase} performs worst since it has the smallest knowledge about the inverse problem. 
The solver~\eqref{eq:proposed_formulation2a}, which solves only for $\vec x$, performs better than the two previous versions.

So far, the magnitude-only solvers, including the versions with partially coherent observations, have shown a rather good convergence rate\,---\,with a slight advantage of~\eqref{eq:proposed_formulation2} over~\eqref{eq:proposed_formulation1} and a great advantage for~\eqref{eq:proposed_formulation2a}. 
However, for increasing $M$, local minima possibly prevent a $100$\si{\percent} reconstruction rate. 
In Fig.~\ref{fig:success1}(d), the \ac{svd} is employed to identify the vector for the smallest singular value in~\eqref{eq:gen_EV_prblm2}. As expected, the transition from failed to successful reconstruction occurs at around 
\begin{equation}
\frac{CM}{N}\ge \frac{C(N-1)}{N(C-1)}\,,
\end{equation}
which is a recasted version of~\eqref{eq:SVD_condition}.
The \ac{svd}-based solver~\eqref{eq:nullspace_phases} performs marginally better in the transition to a certain reconstruction in Fig.~\ref{fig:success1}(e).

The minimizations according to~\eqref{eq:proposed_formulation1} and~\eqref{eq:proposed_formulation2} never reach a certain reconstruction since they strongly depend on the initial guess, and sometimes the spectral method fails in this respect.

At an oversampling ratio where the minimizers come close to a success rate of $100$\si{\percent}, the SVD null-space solutions  are able to reach a certain reconstruction.
The comparison method~\eqref{eq:PAULUS} is non-convex without guaranteed convergence and may get stuck in local minima, but empirically it performs here quite well. 
The downsides of this method become apparent later.
Also, the minimization of~\eqref{eq:proposed_formulation2a} achieves certain reconstruction in the  hitherto presented results.

\begin{figure}[!tp]
\centering
\includegraphics{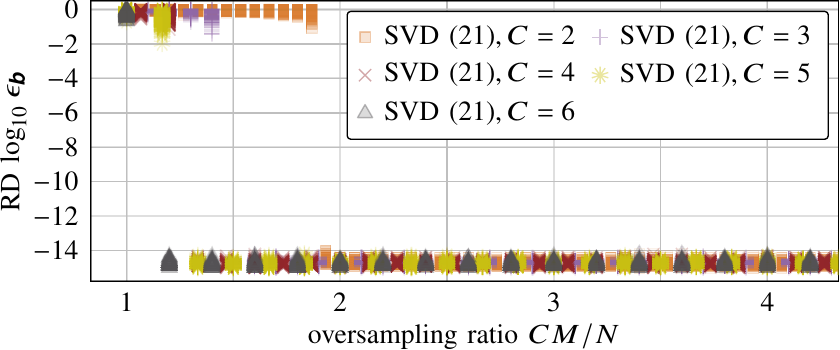}
\caption{RD of the inverse problem solved for Gaussian random matrices, $N=30$, $n=0$. Each $CM/N$ ratio with $200$ simulations.\label{fig:scatter-noisefree}}
\end{figure}

Finally, we investigate the noise-free case, i.e., $n=0$, for the second null-space solver~\eqref{eq:nullspace_phases} in Fig.~\ref{fig:scatter-noisefree}.
The achievable accuracy is below $10^{-14}$ once the necessary oversampling criteria are met, e.g., at $2M/N=58/30$ for $C=2$.
All proposed phase retrieval algorithms gain reconstruction accuracy once the ideal noise-free case is considered. 
The accuracy of the cost-function minimizations of course depends on the stopping criteria for the iterative solver.

\subsection{A Larger Scenario with $N=3000$}
\begin{figure}[!t]
\centering
\subfloat[]{\includegraphics{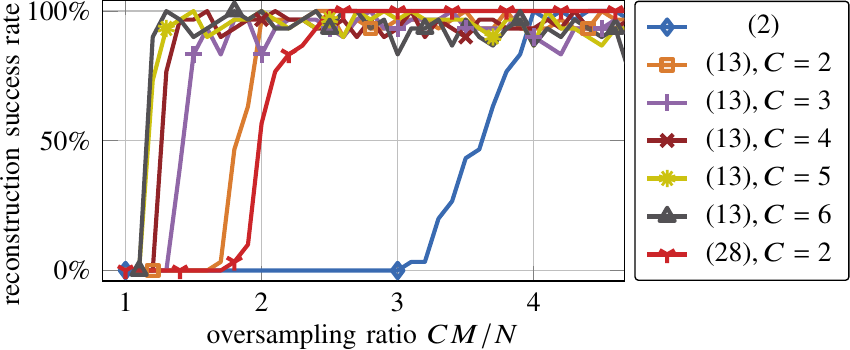}}\\
\subfloat[]{\includegraphics{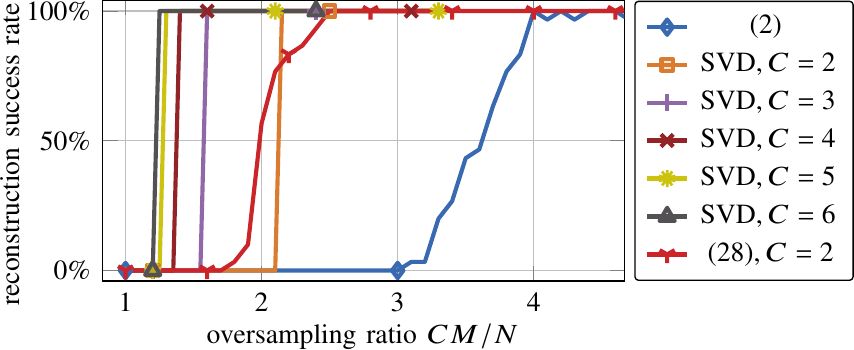}}
\caption{Success rate for Gaussian matrices, $N=3000$ and $n=10^{-2}$, over $CM/N$. (a) Non-linear cost function solver~\eqref{eq:proposed_formulation2a}. (b) SVD-solver~\eqref{eq:gen_EV_prblm2}.\label{fig:success5}}
\end{figure}

Now, the noise level is set to $n=10^{-2}$ and the number of unknowns is increased to $N=3000$.
Among the non-linear minimization techniques, we investigate only the hitherto best one, which is~\eqref{eq:proposed_formulation2a}. The two SVD-based solvers were on par so far and we investigate just one of them.
The success rates, again for a threshold of $3n$, are depicted for two solvers in Fig.~\ref{fig:success5}.
Two differences to the $N=30$ case are observed. 

On the one hand, the SVD-based solver requires a slightly larger oversampling ratio for convergence than expected from the threshold according to~\eqref{eq:SVD_condition}. E.g., for $C=2$, the threshold for a successful reconstruction is at $CM/N=1.996$, but success is only observed at $CM/N>2$. The reason is the interaction of a noise-induced transition period, compare Fig.~\ref{fig:scatter_split_2} to Fig.~\ref{fig:scatter-noisefree}, and the demanding success threshold of $3n$.

On the other hand, the cost-function minimization~\eqref{eq:proposed_formulation2a} shows a worse success rate than for the case $N=30$. 
In particular, it suffers from local minima for this example; the same happens for the noise-free case and large $N$.
If our goal is a certain reconstruction rate of $100$\si{\percent}, only the linear solver in Fig.~\ref{fig:success5} beats the magnitude-only solver and the state-of-the-art solver~\eqref{eq:PAULUS}. 
It becomes clear that the non-linear non-convex minimizations fail to ensure  a correct reconstruction. The SVD-based solvers are able to provide absolute certainty at the theoretically determined thresholds dependent on $C$, within a noise-caused margin over the expected threshold.

\begin{figure}[!t]
\centering
\subfloat[]{\includegraphics[]{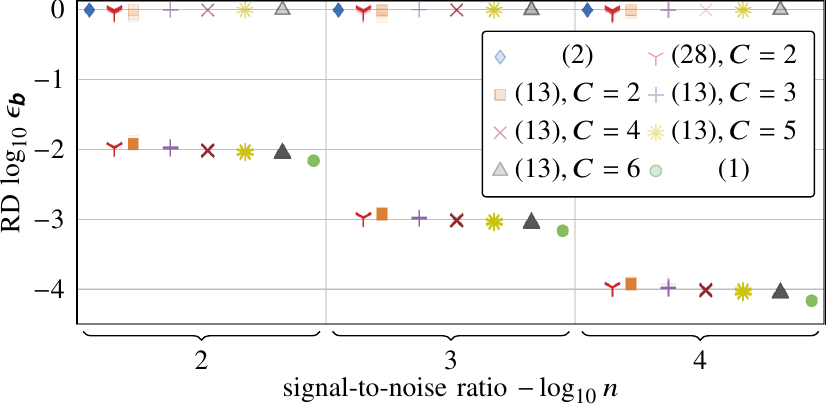}}\\
\subfloat[]{\includegraphics[]{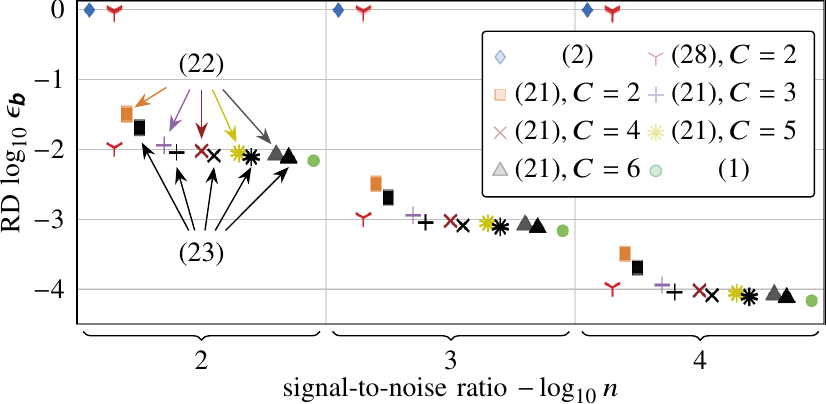}}
\caption{Signal-to-noise ratio analysis, 100 random simulations, $N=3000$, $CM=2.1N$, $n\in\{10^{-2},10^{-3},10^{-4}\}$. 
(a)~Non-linear solver~\eqref{eq:proposed_formulation2a}.
(b)~Linear solver~\eqref{eq:nullspace_phases} with phase reconstruction according to~\eqref{eq:rec1} and~\eqref{eq:rec2}.
\label{fig:noise}}
\end{figure}

This influence of noise on the behavior of the non-linear solver~\eqref{eq:proposed_formulation2a} and the linear solver~\eqref{eq:nullspace_phases} is analyzed in Fig.~\ref{fig:noise} 
for $N = 3000$, $CM=2.1N$, $100$ random simulations, and $n\in\{10^{-2},10^{-3},10^{-4}\}$.
Comparison methods comprise the standard complex (best-performing) and phaseless (worst-performing) solvers and the multi-probe comparison algorithm for $C=2$ (often suffering from local minima).
The performance of the partially-coherent phaseless solvers for $C\inN{[2,6]}$ depends linearly on the chosen signal-to-noise ratio,  except for local minima. 
The linear method~\eqref{eq:nullspace_phases} with enforced magnitude-one constraints~\eqref{eq:rec2} reliably provides a good solution.

\section{An Exemplary Application: Phase Retrieval for Synthetic Antenna Near-Field Measurement Data\label{sec:res_nffft}}
\begin{figure}[t]
\centering
\includegraphics[]{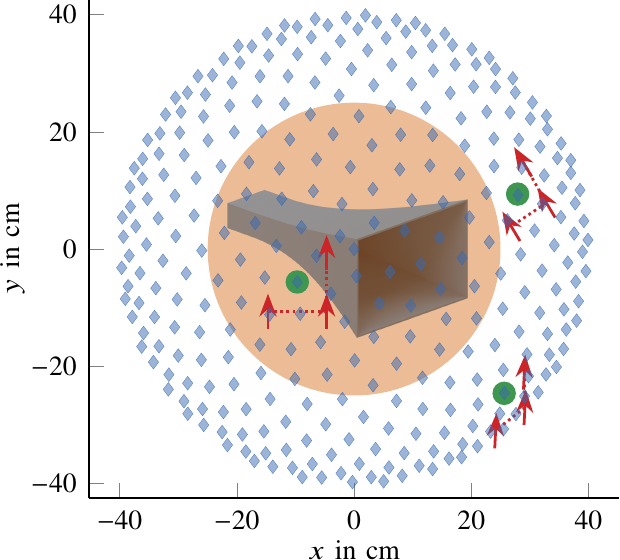}
\caption{Synthetic antenna near-field measurement setup. An L-shaped probe array (red arrows) around the measurement reference position (green dot) is used to acquire magnitude and local phase difference information (here: $C = 3$) at the sample locations (blue diamonds). Probe locations (red arrows) at different sampling locations do not coincide; phase concatenation is not possible. The solution vector $\vec x$ corresponds to surface currents densities placed tangentially on the smallest sphere (orange sphere) enclosing the horn antenna.\label{fig:horn_setup}}
\end{figure}

We now consider a synthetic antenna near-field measurement setup as illustrated in Fig.~\ref{fig:horn_setup}. 
For more information on the idea of \acp{nffft}, for instance refer to~\cite{ludwig_nearfield_1971,hansen_spherical_1988,Sarkar.1999,alvarez_reconstruction_2007,Schmidt.2008,qureshi_efficient_2013,quijano2010field,kornprobst2019solution}. 
As part of a phaseless \ac{nffft}, the phases of the observed \ac{nf}s are to be reconstructed. 
In the context of the setup in Fig.~\ref{fig:horn_setup}, the measurement vector $|\vec b|$ corresponds to the signals received by known probe antennas placed at sample locations (blue diamonds) on a closed hull surrounding the \ac{aut}, which here is a horn antenna. 
In order to model the electromagnetic radiation of the \ac{aut}, the unknown coefficients $\vec x$ of equivalent sources on an enclosing surface (orange sphere) are introduced. 
The received probe signals and the coefficients of the equivalent sources are linked via the electromagnetic radiation operator $\mat A$, which typically does not feature Gaussian distributed rows. 
Skipping the details\,---\,we refer the interested reader to~\cite{Bleistein1977,quijano2010field,kornprobst2019solution,8603822}\,---,\,the spectrum of $\mat A$ is strongly decaying and typically exhibits a non-trivial kernel, just to name two major differences to Gaussian matrices.

Assuming a probe antenna array and coherence between the probe elements, we are able to apply the presented phase retrieval algorithms. 
To pick reasonable multi-antenna probes, we recapitulate that this measurement setup and its field distributions are three-dimensional, but a two-dimensional description on the measurement surface is sufficient. 
Linking phases in two dimensions is possible in the case of a three-element probe array with an L-shape: 
Two linearly independent phase differences are acquired at every measurement location, resulting in $C = 3$. 
The L-probe (red arrows) placed at an exemplary measurement location (green dot) is illustrated in Fig.~\ref{fig:horn_setup}. 
For a comparison case with $C=2$, we pick the two diagonal probe elements only.
Real-world measurement setups for such a scenario are found in~\cite{Costanzo.2001,Costanzo.2001b,Costanzo.2005,Costanzo.2008,Paulus.2017b,Sanchez.2020}. We stick to synthetic data since i) only simulated results offer the knowledge of the true solution and ii) studies on a large set of different antennas are not really feasible with measurements.

In the considered synthetic measurement setup, the equivalent-source sphere enclosing the \ac{aut} and the measurement sphere exhibit diameters of $5$ and $8$ wavelengths, respectively. As equivalent sources, $N = 1200$ tangential Hertzian dipoles are utilized and the horizontal as well as the vertical spacing between the probe-array elements is one wavelength. Each probe element is modeled as a single Hertzian dipole.

\begin{figure}[t]
\centering
  \subfloat[]{\includegraphics{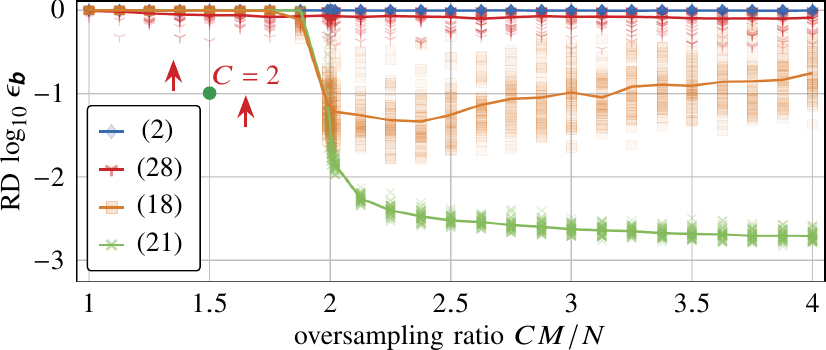}}\\
  \subfloat[]{\includegraphics{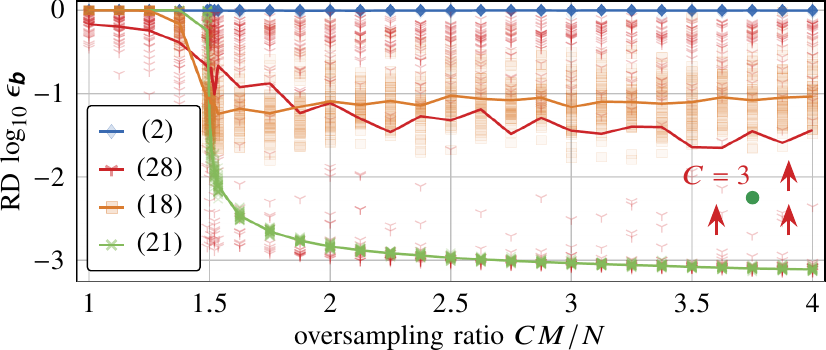}}
\caption{RD of the inverse problem solved for a synthetic antenna near-field measurement setup. 
The average RDs are indicated by solid lines. (a)~Utilizing a probe with two diagonal array elements ($C = 2$). (b)~Employing a three-element, L-shaped probe array ($C = 3$).\label{fig:horn_results}}
\end{figure}

The obtainable RDs for the two known formulations~\eqref{eq:std_wo_phase} and~\eqref{eq:PAULUS}, as well as the two proposed ones~\eqref{eq:nullspace_phases} and~\eqref{eq:gen_EV_prblm2}, is depicted in Fig.\,\ref{fig:horn_results}. 
The described cases of $C=2$ and $C=3$ are given in Fig.\,\ref{fig:horn_results}(a) and (b), respectively.
For every ratio of $CM/N$, $50$ random orientations of the \ac{aut} were simulated, resulting in different measurement vectors. 
All results were obtained for a noise-to-signal ratio of $n = 10^{-3}$.

\begin{figure}[t]
\centering
 \subfloat[]{\includegraphics{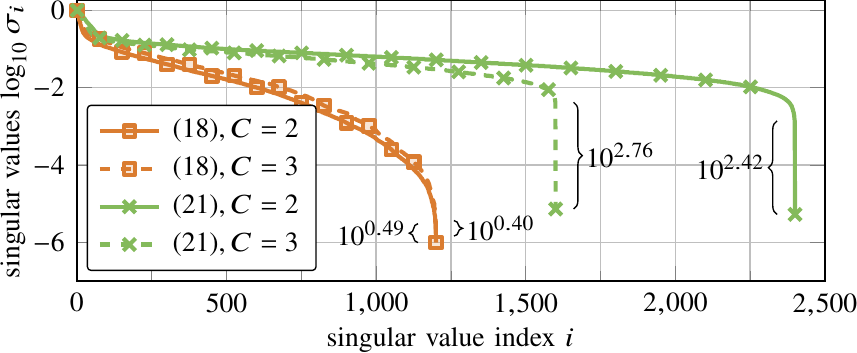}}\\
 \subfloat[]{\includegraphics{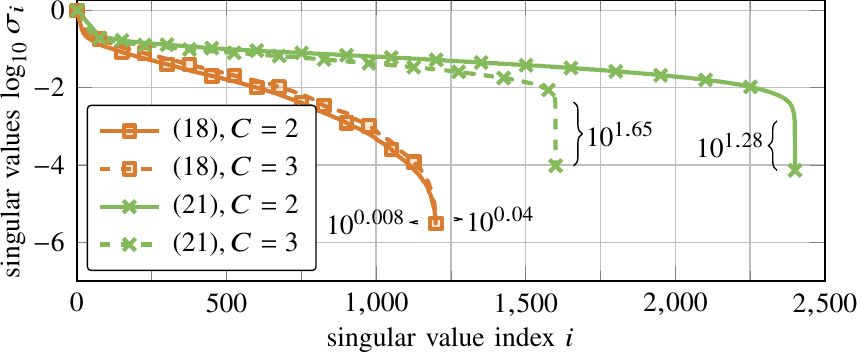}}
\caption{Normalized SVD spectra for $\mat Q$ of~\eqref{eq:gen_EV_prblm2} and $\mat R$ of~\eqref{eq:nullspace_phases} for the cases $C=2$ and $C=3$ of the synthetic measurement data. 
(a)~Noise-free case $n=0$. 
(b)~Noise-contaminated observations with $n=10^{-3}$.\label{fig:horn_svds}}
\end{figure}

In both cases, the best results are obtained with formulation~\eqref{eq:nullspace_phases}, which is observed to reliably yield accurate results above a certain ratio of $CM/N$. 
The existing formulations~\eqref{eq:std_wo_phase} and~\eqref{eq:PAULUS} are observed to either fail completely or they are not guaranteed to find a satisfactory solution. 
All formulations exploiting the phase differences are observed to yield better results for the case of the full L-shaped probe ($C = 3$) compared to the two-element diagonal probe ($C = 2$). 
Especially formulation~\eqref{eq:PAULUS} is observed to benefit significantly. 
It yields similar results as~\eqref{eq:gen_EV_prblm2} in the case of $C = 3$. 
The difference between the two proposed null-space formulations,~\eqref{eq:gen_EV_prblm2} and~\eqref{eq:nullspace_phases}, can be explained by looking at the spectrum of the singular values of $\mat Q$ and $\mat R$. 
Considering a noise-free setup with $CM/N = 3$ and the L-shaped probe, the spectra of $\mat Q$ and $\mat R$ are given in Fig.~\ref{fig:horn_svds}. 
The null-space is more distinct for~\eqref{eq:nullspace_phases}, i.e., the ratio of the second smallest singular value to the smallest one is significantly greater for $\mat R$ than for $\mat Q$. 

Whenever perturbations affect the observations, and thus the spectrum of the singular values, it is more difficult to avoid false solutions and to maintain the desired null-space of $\mat Q$. 
Since $\mat R$ features a more pronounced separation between its non-trivial null-space vector (smallest singular value) and false solutions (any other singular value), this formulation is more robust with respect to noise for the considered exemplary scenario and its forward operator.

\section{Conclusion}
Various formulations of the phase retrieval problem with additional knowledge of phase differences within subsets of observations have been presented. 
Two of the formulations are based on homogeneous linear systems of equations and require only defined oversampling and mild conditions to work with certainty. 
Since the non-linear magnitude constraints are only \mbox{implicitly} fulfilled in these linear equations, they do not exploit the full available information and may be seen as \emph{suboptimal yet simply reachable} solutions.
In order to improve the solution obtained in particular for cases with insufficient oversampling, a minimization problem for a non-linear cost functional has also been presented, which provides better results than comparable algorithms found in literature.

The presented results demonstrate that reliable phase retrieval for partially coherent observations is possible if certain conditions on the phase retrieval equations are fulfilled.
In the case of a well-conditioned random Gaussian forward operator, the performance of the two approaches is very similar.
However, significant performance advantages are found for the phase reconstruction with projection-based method~\eqref{eq:nullspace_phases}, in particular in the presence of noise. 
Our explanation is that the projection matrix~\eqref{eq:nullspace_phases} removes the influence of the conditioning of the forward operator (i.e., its decaying singular value spectrum).

\bibliographystyle{IEEEtran}
\bibliography{IEEEabrv,phaseless}
\vspace*{-1cm}
\begin{IEEEbiography}
 [{\includegraphics[width=1in,height=1.25in,clip,keepaspectratio]{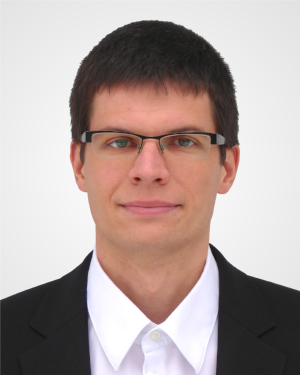}}]
{Jonas Kornprobst} (Graduate Student Member, IEEE) received the B.Eng. degree in electrical engineering and information technology from the University of Applied Sciences Rosenheim, Rosenheim, Germany, in 2014, and the M.Sc. degree in electrical engineering and information technology from the Technical University of	Munich, Munich, Germany, in 2016. 
Since 2016, he has been a Research Assistant with the Chair of High-Frequency Engineering, Department of Electrical and Computer Engineering, Technical University of Munich. 
His current research interests include numerical electromagnetics, in particular integral equation	methods, antenna measurement techniques, antenna and antenna array design, as well as microwave circuits.
\end{IEEEbiography}
\vspace*{-1cm}

\begin{IEEEbiography}
 [{\includegraphics[width=1in,height=1.25in,clip,keepaspectratio]{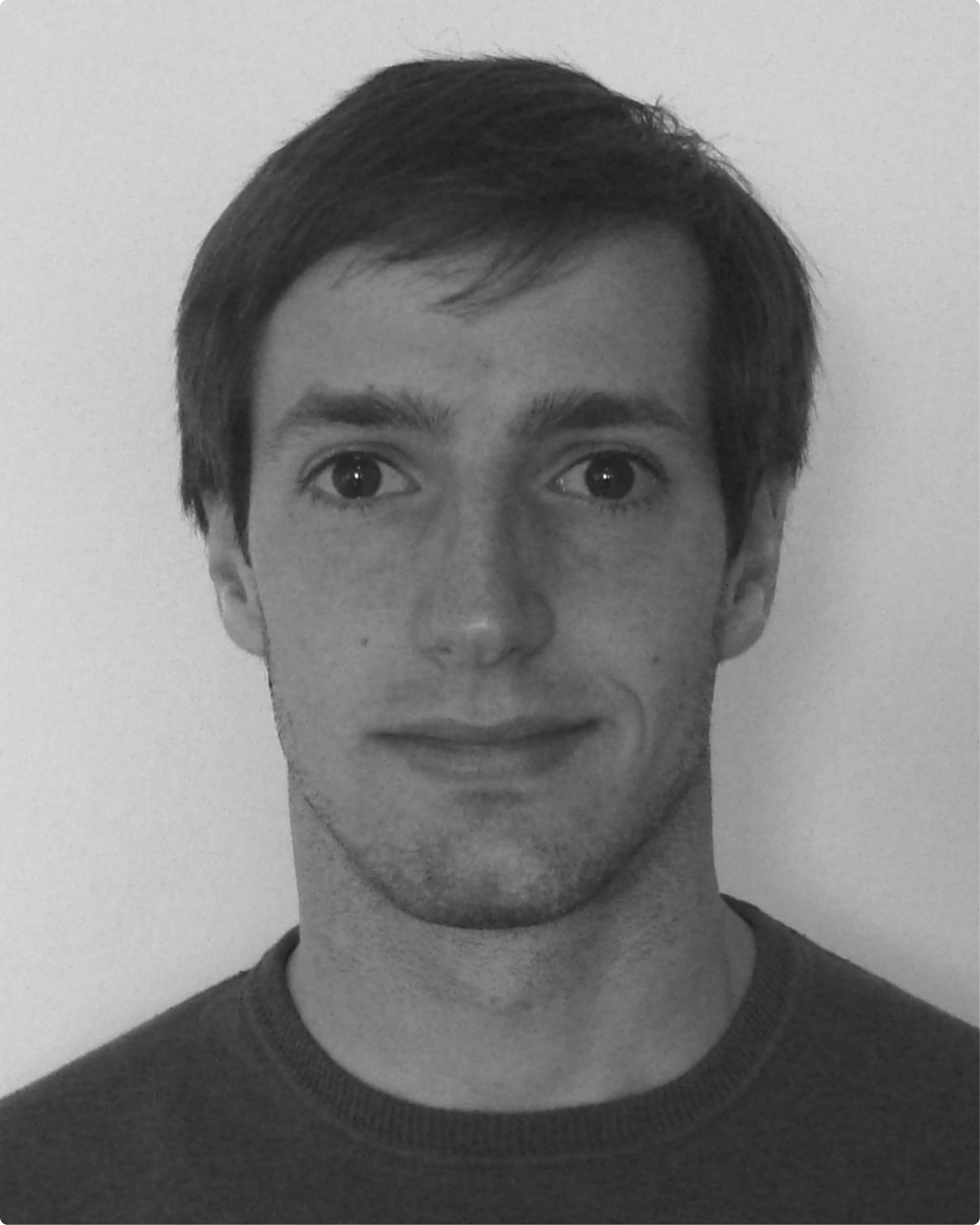}}]
{Alexander Paulus} (Graduate Student Member, IEEE) received the M.Sc. degree in electrical engineering and information technology from the Technical University of Munich, Munich, Germany, in 2015. 
Since 2015, he has been a Research Assistant at the Chair of High-Frequency Engineering, Department of Electrical and Computer
Engineering, Technical University of Munich. 
His research interests include inverse electromagnetic problems, computational electromagnetics and antenna measurement techniques.
\end{IEEEbiography}
\vspace*{-1cm}
\begin{IEEEbiography}
 [{\includegraphics[width=1in,height=1.25in,clip,keepaspectratio]{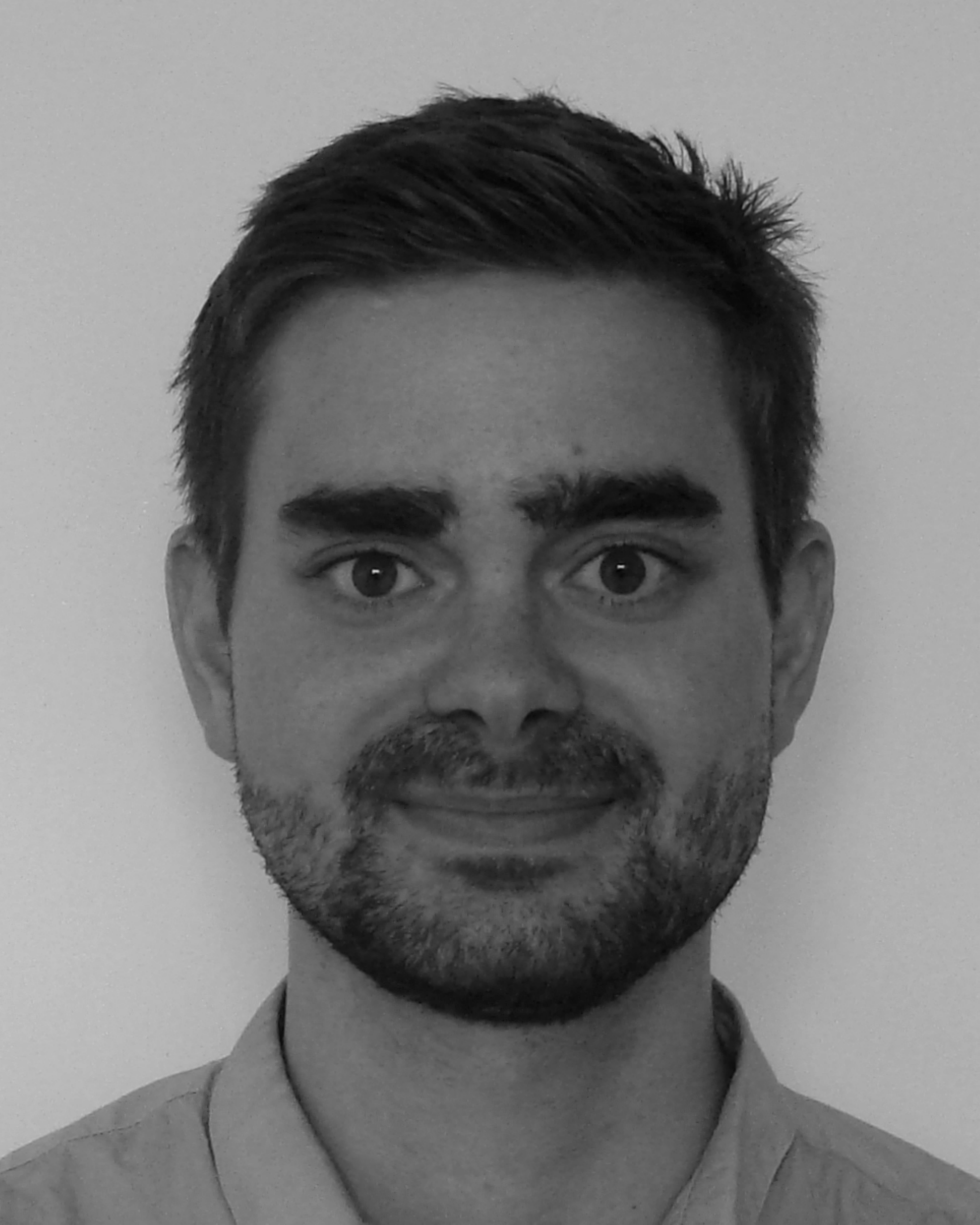}}]
{Josef Knapp} (Graduate Student Member, IEEE) received the M.Sc. degree in electrical engineering and information technology from the Technical University of Munich, Munich, Germany, in 2016. 
Since 2016, he has been a Research Assistant at the Chair of High-Frequency Engineering, Department of Electrical and Computer
Engineering, Technical University of Munich. 
His research interests include inverse electromagnetic problems, computational electromagnetics, antenna measurement techniques in unusual environments, and field transformation techniques.
\end{IEEEbiography}
\vspace*{-1cm}
\begin{IEEEbiography}
 [{\includegraphics[width=1in,height=1.25in,clip,keepaspectratio]{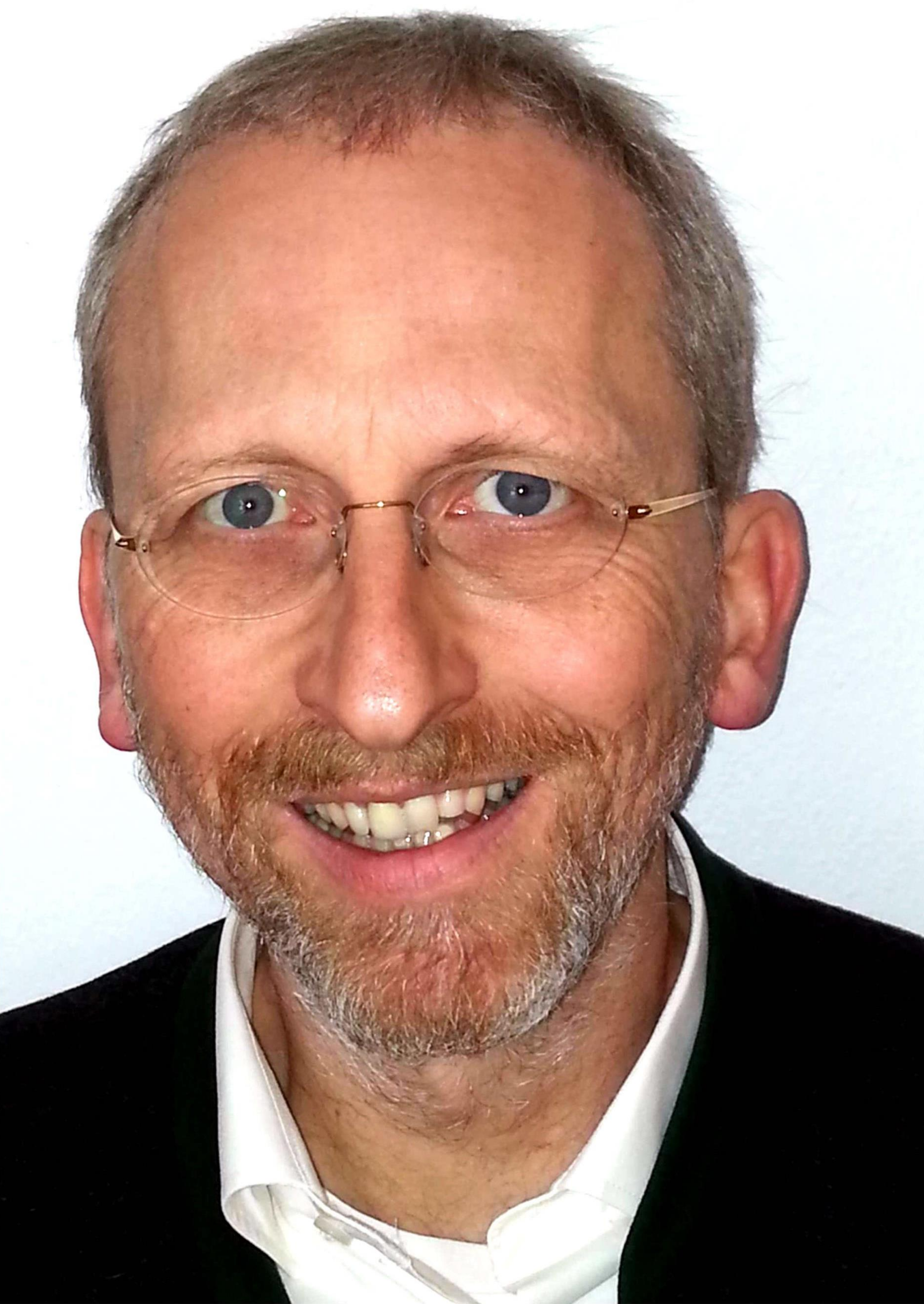}}]
{Thomas F.~Eibert} (Senior Member, IEEE) received the Dipl.-Ing.\,\,(FH) degree from Fachhochschule N\"urnberg, Nuremberg, Germany, the Dipl.-Ing.~degree from Ruhr-Universit\"at Bochum, Bochum, Germany, and the Dr.-Ing.~degree from Bergische Universit\"at Wuppertal, Wuppertal, Germany, in 1989, 1992, and 1997, all in electrical engineering. 
He is currently a Full Professor of high-frequency engineering at the Technical University of Munich, Munich, Germany. 
His current research interests include numerical electromagnetics, wave propagation, measurement and field transformation techniques for antennas and scattering as well as all kinds of antenna and microwave circuit technologies for sensors and communications.		
\end{IEEEbiography}

\end{document}